\documentclass[a4paper,11pt]{article}
\usepackage[english]{babel}
\usepackage[super,comma,sort&compress]{natbib}
\usepackage{url}
\usepackage[noblocks]{authblk}
\makeatletter
\renewcommand\@biblabel[1]{#1} 
\makeatother
\usepackage{latexsym,amssymb,amsmath,amsthm}
\usepackage[dvips]{graphicx}
\usepackage{color}

\begin{document}
\title{Assessing the distribution of discrete survival time in presence of recall error}

\author[]{Sedigheh Mirzaei Salehabadi, Edwina Yeung, Germaine M. Buck Louis, Rajeshwari Sundaram}
\affil[]{{\small {\it Eunice Kennedy Shriver} National Institute of Child Health and Human Development, Bethesda, USA.}}
 \date{}
\maketitle

\begin{abstract}
Retrospectively ascertained survival time may be subject to recall error. An example of discrete survival time with such recall error is time-to-pregnancy (TTP), the number of months non-contracepting couples require to get pregnant which is a measure of human fecundity. The epidemiological literature has demonstrated that retrospective TTP is subject to recall error and statistical models focusing on TTP have not accounted for the recall error. We propose a multistage model that utilizes women's retrospectively-reported TTP and associated certainty to estimate the TTP distribution. Our proposed model utilizes a discrete survival function that accounts for random heterogeneity arising from between women TTP data as well as a multinomial regression model to account for her certainty as accuracy may decline over time, i.e., depends on time since pregnancy in estimating the TTP distribution. Other novel features of the model include attention to whether the pregnancy was (un)planned as well as providing an approach to predict survival function for women without a reported TTP. Our model allows for the consideration of covariates for each of the underlying factors of (un)planned pregnancy, measure of certainty and TTP distribution. The proposed model is applicable for any discrete survival time when certainty in reporting may be a consideration. We use Monte Carlo simulations to assess the finite sample performance for the proposed estimators. We illustrate our proposed method using data from Upstate KIDS Study.\\

\noindent\emph{Discrete survival; Random effect; Retrospective study; Multinomial logistic regression.}
\end{abstract}

\newpage

\section{Introduction}
\label{intro}
\vspace*{-2mm}
Retrospectively ascertained time to event is often encountered in observational studies in both biomedical and social sciences literature. Time-to-pregnancy (TTP) is an example of time to event data. Retrospectively ascertained time-to-event observations are typically subject to recall error as memory may decline over time. Some other instances of recalled time-to-event data subject to recalled error are age at puberty, age at menarche \citep{b14}, duration of breastfeeding or time-to-weaning \citep{b7} and time-to-return of ovulation and menstruation following delivery. One is often interested in the
probability distribution of the underlying time to event, as it is useful for comparing populations, setting benchmarks for individuals and so on. Hence, estimation of the distribution function of retrospectively observed time-to-event with recall error is an important inferential problem.

Human fecundity is the biologic capacity of men and women for reproductive irrespective of their pregnancy intentions. TTP, the length of time non-contracepting couples require to become pregnant an it is an important measure of human fecundity. TTP is discrete survival time as couples can get pregnant only once within a menstrual cycle. The statistical literature has devoted considerable attention to building meaningful models for TTP. For prospectively ascertained TTP, Beta-Geometric distribution for modeling TTP was proposed by Weinberg and Gladen (1986)\cite{b20}, and later Skinner and Humphreys (1997)\cite{b17}, proposed a discrete survival model which also incorporated couple-specific heterogeneity. Subsequently Sundaram et~al. (2012)\cite{b19} proposed a discrete survival model that extended Scheike and Jensen's model by incorporating the couple's intercourse behavior during the fertile window.
 Additional models in case of TTP ascertained from current duration design have been studied by Keiding et~al. (2002)\cite{b9} and McLain et~al. (2014)\cite{b11}. However, all the above mentioned models were used for prospectively ascertained TTP with no uncertainty in time-to-event information. To the best of our knowledge, there has been no estimation of retrospectively ascertained TTP with recall bias.

Epidemiological studies have assessed the validity of retrospectively recalled TTP, and reported a negative association with length of recall \citep{b21,b4,b8}. In light of increasing research that utilizes TTP as both an outcome in relation to various covariates or as a predictor of infant and maternal health outcomes, statistical methods are needed to account for study design and reliance on retrospective TTP. Thus, a model for discrete survival time that accounts for certainty attached to the retrospectively ascertained information as well, as other unique features which are typically encountered in TTP studies, is needed.

The Upstate KIDS Study \citep{b2} collected TTP data from mothers and provided us with the motivation to propose a multistage model that incorporate information about pregnancy intentions as well as level of certainty in self-reported TTP. Briefly, this is a population-based prospective cohort study of 6171 infants (singletons and multiples) born to 5034 women in Upstate New York (excluding the 5 boroughs of New York City which has its own live birth registry) between July 2008 and May 2010. Briefly, mothers were queried about their reproductive histories with particular attention to pregnancy intentions and TTP when her child was about 36 months old. Also, mothers were queried about pregnancy planning and their level of certainty (`very sure', `somewhat sure', `a little sure', `not at all sure') about self-reported TTP. We sought to model how certainty affects the TTP distribution in light of research demonstrating declining accuracy with time \citep{b10,b13,b12}. Motivated by this, we model the recall TTP based on it in the next section.
Even though there are a number of approaches from which the distribution of the TTP may be estimated, there is no suitable model and method for data complicated by two factors: 1) women with unplanned pregnancies who have no well-defined TTP and more importantly 2) women's recalled TTP certainty varies. Such data is not bounded to TTP, it can also arise in other studies of retrospectively collected discrete survival times with indicators of recall uncertainty, as well as sub-strata where the time-to-event is not well-defined.

In this paper, we propose a new approach for estimating the distribution of TTP that incorporates the recall information as well as planning intentions in Section~\ref{model}. In this model, the time of observation is assumed to be independent of the TTP, a logistic regression model is proposed to model planning intention for pregnancy and a multinomial regression model is used to model the different levels of uncertainty. Section~\ref{prediction} details the prediction approach, using a Monte Carlo empirical Bayes technique to predict survival function of TTP given covariates. Results of Monte Carlo simulations are reported in Section~\ref{simulation}. We present detailed analyses of the Upstate KIDS study, undertaken by the National Institutes of Health pertaining to estimation of survival function of TTP in Section~\ref{dataAnalysis}. The simulation studies presented are in line with this application. Some concluding remarks are provided in Section~\ref{concluding}.

\section{Multistage Model for Retrospective Survival Time}
\label{model}
Let $T_i$ denote the time-to-pregnancy for the $i$-th responder with distribution $F_i$ $(i= 1,\ldots,n).$ Let $O_1,\ldots,O_n$ denote the observation times for the $n$ responders. It is assumed that the $O_i$'s are samples from another distribution and are independent of $T_i$'s.
Suppose $\delta_i$ indicates `having certainty answer' and $\xi_i$ indicates `planned pregnancy' for the $i$-th responder with $p_i=Pr(\xi_i=1)$ to be the probability of planned pregnancy for the $i$-th responder. Let $Z_i$ be the {\it r}-dimensional vector of covariates for the $i$-th respondent, assumed to be independent of $O_i$. Note that the distribution of $T_i$ would depend on $Z_i$.
We define random variable $\varepsilon_i$ as follows to indicate how certain the $i$-th responder is about her answer.
\begin{equation}
\varepsilon_i= \left\{
\begin{array}{ll}
1 & \text {if respondent is very sure about her answer},\\
2 & \text {if respondent is somewhat sure about her answer},\\
3 & \text {if respondent is a little sure about her answer},\\
4 & \text {if respondent is not at all sure about her answer}.
\end{array} \right. \label{epsilon}
 \end{equation}
We regard the above multiple possibilities as outcomes of a multinomial selection.
Thus, we model the allocation probabilities as follows
 allowing their probabilities to depend on the observation time, the time-to-pregnancy and $Z_i$.
\begin{equation}
P(\varepsilon_i=k|O_i=o,T_i=t,Z_i=z)=\pi_{\eta}^{(k)}(k,o,t,z), \quad k=1,2,3,4.
\label{pifunc}
\end{equation}
where $\sum_{k=1}^{4}\pi_{\eta}^{(k)}(k,o,t,z)=1$, and $\eta$ is a vector of parameters.

We refer to the set-up described in the first paragraph of this section, together with \eqref{epsilon} and \eqref{pifunc} as the proposed model. According to this model, there would be five cases for an individual $i$ (irrespective of whether the pregnancy is planned or unplanned), with different contributions to the likelihood.
\begin{description}
 \item [{\sc Case}](i) When $\delta_i=0$ (the $i$-th individual didn't provide how sure she is about her answer), the contribution
of the $i$-th individual to the likelihood is $\sum_{k=1}^{4}P(T_i=t)\pi_{\eta}^{(k)}(k,,o,t,z)$.
\item [{\sc Case}] (ii): When $\delta_i=1$ and $\varepsilon_i=1$ (the $i$-th individual is very sure about her answer), the
contribution of the individual to the likelihood is
$P(T_i=t)\pi_{\eta}^{(1)}(1,o,t,z)$.
\item[{\sc Case}] (iii): When $\delta_i=1$ and $\varepsilon_i=2$ (the $i$-th individual is somewhat sure about her answer), the contribution of the individual to the likelihood is $P(T_i=t)\pi_{\eta}^{(2)}(2,o,t,z)$.
\item[{\sc Case}] (iv): When $\delta_i=1$ and $\varepsilon_i=3$ (the $i$-th individual is a little sure about her answer), the contribution of the individual to the likelihood is $P(T_i=t)\pi_{\eta}^{(3)}(3,o,t,z)$.
\item[{\sc Case}] (v): When $\delta_i=1$ and $\varepsilon_i=4$ (the $i$-th individual is not at all sure about her answer), the contribution of the individual to the likelihood is $P(T_i=t)\pi_{\eta}^{(4)}(4,o,t,z)$.
\end{description}

\begin{figure}[]
\centering
 \includegraphics[height=4.5in,width=3.5in,angle=-90]{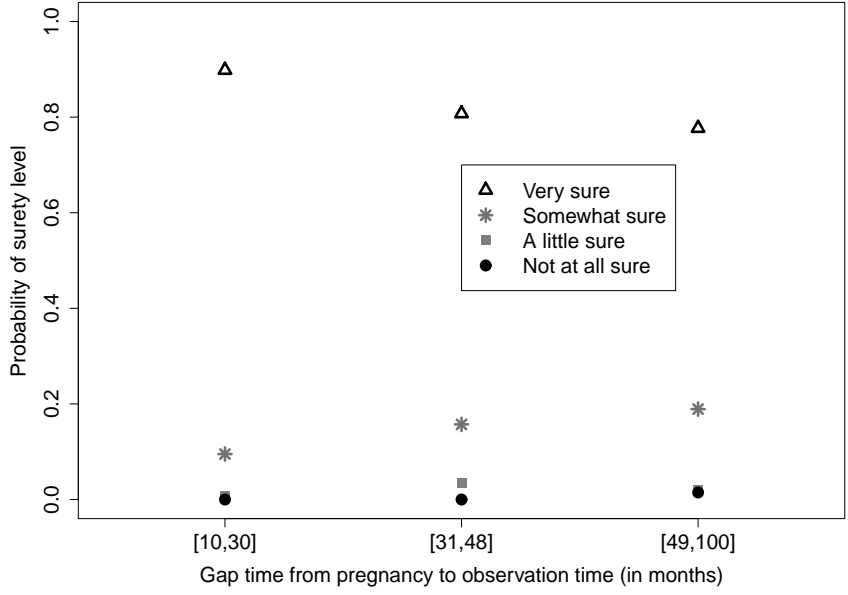}
\caption[]{Probability of certainty for different ranges of gap time from pregnancy to observation time.} \label{fig_1}
\end{figure}

In our motivating study, we have an additional issue of oversampling of pregnancies conceived through infertility treatment. In particular, spontaneous pregnancies versus infertility treated pregnancies were sampled in the ratio of 3:1. We account for this by using sampling weights $w_i, i= 1, \cdots, n$ based on proportions derived from New York State birth certificate data over the period of recruitment (2008-2010) so that the sample is representative of the underlying population of Upstate New York pregnancies for that time period. Therefore, the overall likelihood can be written as
\small
\begin{align}
\prod_{i=1}^{n}\Biggl(&\Big[\sum_{k=1}^{4}P(T_i=t)\pi_{\eta}^{(k)}(k,o,t,z)\Big]^{1-\delta_i}
\Biggl[\prod_{k=1}^{4}\left(P(T_i=t)\pi_{\eta}^{(k)}(k,o,t,z)\right)^{I_{(\varepsilon_i=k)}} \Biggr]^{\delta_i}\times p_i\Biggr)^{w_i\xi_i} \times \notag \\
&\Biggl(\Big[\sum_{k=1}^{4}P(T_i=t)\pi_{\eta}^{(k)}(k,o,t,z)\Big]^{1-\delta_i}
\Biggl[\prod_{k=1}^{4}\left(P(T_i=t)\pi_{\eta}^{(k)}(k,o,t,z)\right)^{I_{(\varepsilon_i=k)}} \Biggr]^{\delta_i}\times(1- p_i)\Biggr)^{w_i(1-\xi_i)}.
\label{ourM1}
\end{align}
\normalsize
Figure~\ref{fig_1} shows the probability of certainty by different groups of gap time from pregnancy to survey time, for the respondents of the Upstate KIDS Study.
It is seen that the gap time order is preserved. Also for shorter gap time of recall, there is greater precision of surety. This finding indicates that memory fades with time, i.e., two women interviewed at the same time would have different level of surety depending on how far her pregnancy time was from the observation time.

In particular, the certainty probability may depend on the time elapsed since pregnancy, $O_i-T_i$ as well as common determinants for TTP. Thus, we use a multinomial function of $(O-T)$ and $Z$ to model $\pi_{\eta}^{(k)}(k,o,t,z)$, for $k=1,2,3,4$. Note that $O-T$ is gap time between pregnancy and date mother completed the questionnare and the dependence is captured through covariates.

To model TTP, we consider the proportional hazards model for discrete survival time. Under this model, the hazard rate $\lambda_i(t)$ of the individual $i$ waiting time $t$, with covariate vector $Z_i$ is
\begin{align}
&\lambda_i(t)=P(T_i=t|T_i\geq t) \notag\\
&\lambda_i(t;z)=1-\exp(-\exp(\beta^TZ_i)),  \label{coxhazard}
\end{align}
where $\beta$ is the regression coefficient \citep{b17}. Note that
\begin{align}
P(T_i=t)&=\lambda_i(t)\prod_{j=1}^{t-1}(1-\lambda_i(j)) \notag\\
&=\lambda_i(t) \exp\left(-\sum_{j=1}^{t-1}\exp(\beta^TZ_{ij})\right).
\end{align}
We further account for differences in distribution of TTP for planned versus unplanned pregnancy by using different vectors of regression coefficients.


To accounts for unobserved covariates or further biological heterogeneity between individuals a random variable $R_i$ is considered.
Conditional on $R_i$ and the intention of planned pregnancy, one obtains
\begin{align}
P(T_i=t|R_i, Z_i,\xi_i)=&\exp \left(-\sum_{j=1}^{t-1}\exp\left(R_i+\rho(j)+\beta^TZ_iI(\xi_i=1)+\phi^TZ_iI(\xi_i=0)\right)\right) \notag\\ &-\exp\left(-\sum_{j=1}^{t}\exp\left(R_i+\rho(j)+\beta^TZ_iI(\xi_i=1)+\phi^TZ_iI(\xi_i=0)\right)\right). \notag
\end{align}
Similar to Scheike and Jensen (1997)\cite{b17}, the marginal form (conditional on random effect $R_i$, covariates $Z_i$ and planned pregnancy) for the probability of conception at time $t$ can be expressed as
\small
\begin{align}
P(T_i=t|R_i,Z_i,\xi_i)=&\left(\frac{1}{\nu. \exp\left(\sum_{j=1}^{t-1}\exp\left(\rho(j)+\beta^TZ_iI(\xi_i=1)+\phi^TZ_iI(\xi_i=0)\right)\right)+1}\right)^{1/\nu} \notag \\ &-\left(\frac{1}{\nu. \exp\left(\sum_{j=1}^{t}\exp\left(\rho(j)+\beta^TZ_iI(\xi_i=1)+\phi^TZ_iI(\xi_i=0)\right)\right)+1}\right)^{1/\nu}.\label{observedTi}
\end{align}
\normalsize

To model planned pregnancy intention we use Logistic regression as follow.
$$Pr(\xi_i=1)=\left(\frac{exp(\gamma^TZ_i)}{1+exp(\gamma^TZ_i)}\right).$$
Considering all the above mentioned factors, the likelihood can be written as
 \begin{align}
\prod_{i=1}^{n}&\Biggl\{\Big[\sum_{k=1}^{4}P(T_i=t|R_i,Z_i,\xi_i)\pi_{\eta}^{(k)}(k,o,t,z)\Big]^{1-\delta_i} \times \notag\\ &\Biggl[\prod_{k=1}^{4}\left(P(T_i=t|R_i,Z_i,\xi_i)\pi_{\eta}^{(k)}(k,o,t,z)\right)^{I_{(\varepsilon_i=k)}}\Biggr]^{\delta_i}\times \left(\frac{exp(\gamma^TZ_i)}{1+exp(\gamma^TZ_i)}\right)\Biggr\}^{w_i\xi_i} \times \notag\\
  &\Biggl\{\Big[\sum_{k=1}^{4}P(T_i=t|R_i,Z_i,\xi_i)\pi_{\eta}^{(k)}(k,o,t,z)\Big]^{1-\delta_i} \times \notag\\
  &\Biggl[\prod_{k=1}^{4}\left(P(T_i=t|R_i,Z_i,\xi_i)\pi_{\eta}^{(k)}(k,o,t,z)\right)^{I_{(\varepsilon_i=k)}}\Biggr]^{\delta_i}\times
  \left(\frac{1}{1+exp(\gamma^TZ_i)}\right)\Biggr\}^{w_i(1-\xi_i)}.
    \label{ourM}
\end{align}
As we have noted before, $\beta$ and $\phi$ are the regression coefficients for covariate effect on TTP distribution for planned and unplanned pregnancies, respectively, $w_i$ is the sampling weight, $\delta_i$ to indicate `having surety answer' and $\xi_i$ indicates `planned pregnancy' for the $i$-th responder.
We model the certainty probabilities through the multinomial logistic as
$$\log\Big(\pi_{\eta}^{(k)}(k,o,t,z)/\pi_{\eta}^{(1)}(1,o,t,z)\Big)=\alpha_{0k}+\alpha_{1k}(o-t)+\alpha^Tz, \quad \ k=2,3,4.\label{logpifunc}$$
Since $\sum_{k=1}^{4}\pi_{\eta}^{(k)}(k,o,t,z)=1.$

The maximum likelihood estimator (MLE) of $\theta=(\nu, \rho_j, \beta, \phi, \gamma, \eta)^T$ are obtained by maximizing the likelihood \eqref{ourM} by setting the partial derivatives equal to zero. A Newton-Raphson iteration may be used to compute the MLEs.

Note that if one ignores the probability of recall in this model, then the likelihood reduces to
 \begin{equation}
\prod_{i=1}^{n}\Biggl\{P(T_i=t|R_i,Z_i,\xi_i)\frac{exp(\gamma^TZ_i)}{1+exp(\gamma^TZ_i)}\Biggr\}^{w_i\xi_i}\times \Biggl\{P(T_i=t|R_i,Z_i,\xi_i)\frac{1}{1+exp(\gamma^TZ_i)}\Biggr\}^{w_i(1-\xi_i)}
\label{norecall}
\end{equation}
This corresponds to the more common scenario of collecting retrospective information without collecting information on uncertainty associated with the recalled time-to-event.
\section{Prediction}\label{prediction}
  An interesting feature of the model and the estimation approach described in the last section is the ability to predict survival probabilities for a new responder $i.$ We consider the scenario where a responder has provided partial information pertaining to covariates but has not provided her time-to-pregnancy as well as whether her pregnancy was planned or unplanned. We consider two stages for the prediction. In the first stage, we use Logistic regression to get the predicted probabilities for whether pregnancy is planned or not. Thus, we can classify the responders with missing planned/unplanned pregnancy intention. In the second stage, to predict survival distribution of TTP, we assume an estimate $\hat\theta$ of $\theta$ of all the underlying parameters for our model with covariance $\Sigma$, estimated by $\hat\Sigma$ based on complete data. Here, it is more relevant to focus on conditional probability of having time-to-pregnancy $u (>t)$, given $t>0$, that is,
\begin{equation}
\pi_i(u|t)\equiv Pr(T_i>u|T_i>t, {\cal D}_i;\theta) \label{pred}
\end{equation}
where ${\cal D}_i=\{T_i,\delta_i,\xi_i; i=1,2,...,n\}$ denotes the sample on which the model was fitted and on which we base our predictions. Using the empirical Bayes estimate, 
Rizopoulos (2011)\cite{b16} proposed the following approximation of (\ref{pred})
\begin{equation}
\hat\pi_i(u|t)=\frac{S_i(u|\hat\theta)}{S_i(t|\hat\theta)}  \quad u>t.   \label{estpred}
\end{equation}
We use arguments of standard asymptotic Bayesian theory (Cox and Hinkley, 1974)[Section 10.6]\cite{b5}, and assume the sample size $n$ is sufficiently large such that $\{\theta|{\cal D}_i\}$ can be well approximated by multivariate $\mathcal{N}(\hat\theta,\hat \Sigma)$. A Monte Carlo estimate of $\pi_i(u|t)$ can be obtained using the following simulation scheme:
\begin{description}
\item[{\sc Step 1.}] draw $\theta^{(l)}\sim \mathcal{N}(\hat\theta,\hat \Sigma)$.
\item[{\sc Step 2.}] compute $\pi_{i}^{(l)}(u|t)$ using \eqref{estpred} with $\theta^{(l)}$.
\end{description}
Steps 1 and 2 are repeated for $l=1,2,\ldots,L$ times. The realization of $\{\pi_i^{(l)}, l=1,2,\ldots,L\}$ can be used to derive estimates of $\hat\pi_i(u|t)$ such as
$$\hat\pi_i(u|t)=\mbox{median}\{\pi_{i}^{(l)}(u|t), l=1,2,\ldots,L\}$$
or
$$\hat\pi_i(u|t)=L^{-1}\sum_{l=1}^{L}\pi_i^{(l)}(u|t),$$
and compute, standard errors using the sample variance over the Monte Carlo samples.

\section{Simulation Studies} \label{simulation}
We compare the performance of MLE's based on proposed likelihood (\ref{ourM}) utilizing recall information (described here as Proposed Recall MLE) and the likelihood (\ref{norecall}) when the recall probability is ignored (described here as Norecall probability MLE). Computation of MLE's in all the cases are done through numerical optimization of likelihood using the Quasi-Newton method \citep{b15}.

For the purpose of simulation, we generate samples of time-to-pregnancy
from a proportional hazards model with probability mass function
\small
\begin{align}
P(T_i=t&|R_i,Z_i,\xi_i)=\left(\frac{1}{\nu. \exp\left(\sum_{j=1}^{t-1}\exp\left(\rho(j)+\beta^TZ_iI(\xi_i=1)+\phi^TZ_iI(\xi_i=0)\right)\right)+1}\right)^{1/\nu} \notag \\ &-\left(\frac{1}{\nu. \exp\left(\sum_{j=1}^{t}\exp\left(\rho(j)+\beta^TZ_iI(\xi_i=1)+\phi^TZ_iI(\xi_i=0)\right)\right)+1}\right)^{1/\nu}. \notag
\end{align}
\normalsize
where the baseline hazard $\rho(j)=\log(-\log(1-U[0,1]))$ and the random effect variable $\exp(R_i)$ is generated from a gamma distribution with mean 1 and variance $\nu=0.5$. The vector of covariates, $Z=(Z_1,Z_2)$, consists of a binary variable, taking values 1 and 0 with probabilities 0.25 and 0.75, and
a continuous variable having the uniform distribution over the
interval [20,45]. We choose the vector of regression coefficients for the TTP distribution of planned and unplanned pregnancy as
$\beta=(\beta_1,\beta_2)=\phi=(\phi_1,\phi_2)=(-0.05,0.01)$ respectively. Further, we generate `gap time from pregnancy to the interview time', i.e.\, $O-T$ from the discrete uniform distribution over
the set $\{43.5,\ldots,48.5\}$. For the coefficients of Logistic regression model, we choose $\gamma=(\gamma_1,\gamma_2)=(0.04,-0.75)$. These choices are in line with the data analytic example discussed in the next section.

In this simulation study, we consider three levels of certainty `very sure', `somewhat sure' and `not at all sure'. We generate certainty probabilities through the multinomial logistic model as
$$\log\Big(\pi_{\eta}^{(k)}(k,o,t,z)/\pi_{\eta}^{(1)}(1,o,t,z)\Big)=\alpha_{0k}+\alpha_{1k}(o-t)+\alpha^Tz, \quad \ k=2,3.\label{logpifunc}$$
Since $\sum_{k=1}^{3}\pi_{\eta}^{(k)}(k,o,t,z)=1,$ the probabilities can be written as
\begin{equation}
 \begin{array}{l}
\pi_{\eta}^{(1)}(1,o,t,z)=1/\big(1+\sum_{k=2}^{3}e^{\alpha_{0k}+\alpha_{1k}(o-t)+\alpha^Tz}\big),\\
\pi_{\eta}^{(k)}(k,o,t,z)=e^{\alpha_{0k}+\alpha_{1k}(o-t)}/\left(1+\sum_{k=2}^{3}e^{\alpha_{0k}+\alpha_{1k}(o-t)+\alpha^Tz}\right), \quad \ k=2,3,
\end{array} \label{pifuncfinal}
\end{equation}
where $\eta=(\alpha_{02},\alpha_{03},\alpha_{12},\alpha_{13},\alpha_1,\alpha_2)$.

We consider two different Scenarios and choose the parameter of recall probability based on these Scenarios. Scenario 1 considers higher possibility for `very sure' answer by choosing parameter $\eta=(-16,-5,0.3,0.05,0.5,0)$, thus the probabilities of different levels of surety at 46 months gap time from pregnancy to survey will be $\pi_{\eta}^{(1)}(46)=0.848, \pi_{\eta}^{(2)}(46)=0.094, \pi_{\eta}^{(3)}(46)=0.057$. This choice is meant to be in line with the example in the next section. Scenario 2 chooses parameter $\eta=(-9,-9,0.195,0.195,0.5,0)$ to have equal probability for different levels of surety. In this Scenario, the probabilities of different level of surety at 46 months gap time from pregnancy to survey will be equal, i.e.\, $\pi_{\eta}^{(1)}(46)=\pi_{\eta}^{(2)}(46)=\pi_{\eta}^{(3)}(46)=0.333$. Note that, 46 months is the median of gap time from pregnancy to survey, in the example of next section. We run 1000 simulations for sample sizes $n=200$ and 1000.



Tables~\ref{t:tableone} and~\ref{t:tabletwo} show the bias, the standard deviation (Stdev), the mean squared error (MSE) and coverage probability (CP) for the MLE's of the parameter $\nu, \beta=(\beta_1,\beta_2), \phi=(\phi_1,\phi_2), \gamma=(\gamma_1,\gamma_2)$, $\eta=(\alpha_{02},\alpha_{03},\alpha_{12},\alpha_{13},\alpha_1,\alpha_2)$ and the estimated probability of `very sure' answer when $o-t=46$, based on the likelihoods (\ref{ourM}) and (\ref{norecall}), for Scenarios 1 and 2, respectively.

 \begin{table}
\caption{\label{t:tableone}Performance of estimated parameters for Scenario 1, with higher possibility for `very sure' answer.}

 \hskip-.5in
 \centering
    \makebox[\textwidth][c]{
 \scalebox{.8}{
\begin{tabular}{c l c cccc c cccc}
      &    &   & \multicolumn{4}{c}{Proposed Recall MLE} && \multicolumn{4}{c}{Norecall Probability MLE} \\
  \cline{4-12}
  $n$ & \multicolumn{2}{c}{Parameters} & Bias& Stdev& MSE& CP& &Bias& Stdev& MSE& CP \\
         \noalign{\hrule height 1pt}
        & Random effect               & $\nu     $      & --0.831	&0.101	 &0.698	  &0.934  & & --1.006 & 0.1250 & 1.283 & 0.924 \\
        \\
        & Regression Coefficient for   & $\beta_1$      & 0.050	    &0.010	 &0.003	  &0.966  & & 0.0532  & 0.0175 & 0.0031 & 0.935 \\
        &  Planned TTP distribution           &$\beta_2$       & --0.010	&0.0202  &0.0005  &0.949  & &--0.017  & 0.1759 &  0.0312 & 0.855 \\
        \\
        & Regression Coefficient for   & $\phi_1$       & 0.050     &0.011   &0.003   &0.938  & &0.0530  & 0.0181 & 0.0031 & 0.928 \\
        &  Unplanned TTP distribution      &$\phi_2 $       & --0.007	&0.302   &0.090   &0.937  & &0.0464  & 0.0103 &  0.0023 & 0.948 \\
        \\
    200 & Regression Coefficient for    &$\gamma_1$      & --0.820   &0.180	 &0.672	  &0.947  & &--0.631 & 0.0563 & 0.4012 & 0.911 \\
        &  Planned Pregnancy (Yes/No) Model     &$\gamma_2$      & -0.012   &0.181   &0.032	  &0.935  & &0.1590  & 0.0563 &  0.0284 & 0.964 \\
       \\
        &                              & $\alpha_{02}$  & 7.950	    &0.560	 &63.51	  &0.941  & &  --    &   --   &   --    &   --   \\
        &                              &$\alpha_{03}$   & --3.010	&1.561	 &11.49	  &0.933  & &  --    &   --   &   --    &   --   \\
        &Regression Coefficient to model & $\alpha_{12}$  & 4.630	    &1.250	 &22.99	  &0.969  & &  --    &   --   &   --    &   --   \\
        &Recall probability for TTP &$\alpha_{13}$   & --2.070	&2.140	 &8.864	  &0.949  & &  --    &   --   &   --    &   --   \\
        &                              &$\alpha_{1}$   & --0.631    &1.501	 &2.646	  &0.961  & &  --    &   --   &   --    &   --   \\
        &                              &$\alpha_{2}$   &  2.703	    &0.924	 &8.143   &0.928  & &  --    &   --   &   --    &   --   \\
        \\
        &  Probability of `very sure' recall &                & 0.030     &0.127	 &0.015	  &0.963  & &  --    &   --   &   --    &   --   \\
        \\
\noalign{\hrule height 1pt}
\\
        & Random effect               & $\nu     $      & --0.420   &0.104	 &0.557	  &0.947  & &--1.003  & 0.1226 & 1.0213 & 0.935 \\
        \\
        & Regression Coefficient for   & $\beta_1$      & 0.040	    &0.007	 &0.002   &0.951  & &0.0434  & 0.0092 & 0.0020 & 0.911 \\
        &  Planned TTP distribution    &$\beta_2$       & --0.022	&0.010   &0.0005  &0.955  & &--0.012 & 0.1456 & 0.0213  & 0.877 \\
        \\
        & Regression Coefficient for   & $\phi_1$       & 0.040     &0.008	 &0.002	  &0.948  & &0.0464  & 0.0103 & 0.0023 & 0.948 \\
        &  Unplanned TTP distribution  &$\phi_2 $       & --0.031	&0.280	 &0.079   &0.951  & &--0.020 & 0.3804 &  0.1451 & 0.943 \\
        \\
    1000& Regression Coefficient for    &$\gamma_1$      & --0.592	&0.041	 &0.349	  &0.947  & &--0.338 & 0.0874 & 0.1222 & 0.917 \\
        &  Planned Pregnancy (Yes/No) Model     &$\gamma_2$      & 0.190 	&0.031	 &0.037   &0.943  & &0.1515  & 0.0574 &  0.0262 & 0.892 \\
        \\
        &                              & $\alpha_{02}$  & 2.580	    &0.364	 &6.786	  & 0.947 & &  --    &   --   &   --    &   --   \\
        &                              &$\alpha_{03}$   & --1.314	&1.020	 &2.756	  &0.916  & &  --    &   --   &   --    &   --   \\
        & Regression Coefficient to model & $\alpha_{12}$  & 3.020     &1.021	 &10.16  &	0.945 & &  --    &   --   &   --    &   --   \\
        & Recall probability for TTP &$\alpha_{13}$   & --1.162	&1.030	 &2.406	  &0.940  & &  --    &   --   &   --    &   --   \\
        &                              &$\alpha_{1}$   & --0.470	&0.980	 &1.181	  &0.954  & &  --    &   --   &   --    &   --   \\
        &                              &$\alpha_{2}$   & 0.991	    &0.902	 &1.793	  &0.942  & &  --    &   --   &   --    &   --   \\
        \\
        &  Probability of `very sure' recall &                & 0.030	    &0.113	 &0.013	  &0.955  & &  --    &   --   &   --    &   --   \\

\noalign{\hrule height 1pt}
\end{tabular}
}
}

\end{table}

 \vskip-.05in

 \begin{table}
    \caption{\label{t:tabletwo}Performance of estimated parameters for Scenario 2, with equal possibility for all levels of certainty.}
\hskip-.3in
 \centering
    \makebox[\textwidth][c]{
 \scalebox{0.8}{
\begin{tabular}{c l c cccc c cccc}
      &    &   & \multicolumn{4}{c}{Proposed Recall MLE} && \multicolumn{4}{c}{Norecall Probability MLE} \\
  \cline{4-12}
  $n$ & \multicolumn{2}{c}{Parameters} & Bias& Stdev& MSE& CP& &Bias& Stdev& MSE& CP \\
         \noalign{\hrule height 1pt}
        & Random effect               & $\nu     $      & --0.960	&0.121	&0.936	 &0.956  & &--1.088 & 0.1306 & 1.2011 & 0.947 \\
        \\
        & Regression Coefficient for   & $\beta_1$      & 0.051	    &0.021	&0.003	 &0.970  & &0.0542  & 0.0198 & 0.0033 & 0.936 \\
        &  Planned TTP distribution           &$\beta_2$       & 0.021	    &0.042	&0.002	 &0.942  & &0.0241  & 0.2263 &  0.0518 & 0.944 \\
        \\
        & Regression Coefficient for   & $\phi_1$       & 0.052	    &0.021	&0.003	 &0.960  & &0.0552  & 0.0214 & 0.0035  & 0.938 \\
        &  Unplanned TTP distribution         &$\phi_2 $       & 0.093	    &0.741	&0.557	 &0.941  & &0.1436  & 0.9813 &  0.9836 & 0.925 \\
        \\
    200 & Regression Coefficient for    &$\gamma_1$      & --0.620	&0.061	&0.388	 &0.936  & &--0.728 & 0.0297 & 0.5311 & 0.907 \\
        &  Planned Pregnancy (Yes/No) Model     &$\gamma_2$      & 0.163	    &0.060	&0.029   &0.940  & &0.2061  & 0.0297 &  0.0434 & 0.918 \\
       \\
        &                              & $\alpha_{02}$  & 4.901	    &0.520	&24.20	 & 0.930 & &  --    &   --   &   --    &   --   \\
        &                              &$\alpha_{03}$   & --2.810	&1.670	&10.62	 &0.941  & &  --    &   --   &   --    &   --   \\
        & Regression Coefficient to model  & $\alpha_{12}$  & 3.601	    &2.231	&17.932	 &0.939  & &  --    &   --   &   --    &   --   \\
        & Recall probability for TTP       &$\alpha_{13}$   & --2.106	&2.631	&11.326	 &0.940  & &  --    &   --   &   --    &   --   \\
        &                              &$\alpha_{1}$   &-1.611	    &1.751	&5.622	 &0.96   & &  --    &   --   &   --    &   --   \\
        &                              &$\alpha_{2}$   &1.600	    &1.071	&3.705	 &0.934  & &  --    &   --   &   --    &   --   \\
        \\
        &  Probability of `very sure' racall &                & 0.443	    &0.310	&0.289	 &0.970  & &  --    &   --   &   --    &   --   \\
\\
\noalign{\hrule height 1pt}
\\
        & Random effect               & $\nu     $      & --0.470	&0.111	&0.5597	 &0.961  & &--1.005 & 0.1141 & 1.0244 & 0.956 \\
        \\
        & Regression Coefficient for   & $\beta_1$      & 0.040	    &0.008	&0.002	 &0.938  & &0.0463  & 0.0079 & 0.0022 & 0.937 \\
        &  Planned TTP distribution          &$\beta_2$       &--0.021	&0.021	&0.001	 &0.959  & &--0.022 & 0.1256 &  0.0162 & 0.952 \\
        \\
        & Regression Coefficient for   & $\phi_1$       &0.045	    &0.008	&0.002	 &0.958  & &0.0459  & 0.0087 & 0.0021 & 0.948 \\
        &  Unplanned TTP distribution         &$\phi_2 $       &--0.04	    &0.291	&0.085	 &0.937  & &--0.100 & 0.6175 &  0.3915& 0.952 \\
        \\
    1000& Regression Coefficient for    &$\gamma_1$      &--0.59	    &0.024	&0.348   &0.938  & &--0.529  & 0.0558 & 0.2837 & 0.933 \\
        &  Planned Pregnancy (Yes/No) Model    &$\gamma_2$      & 0.192	    &0.031	&0.037	 &0.972  & & 0.1170  & 0.0255 &  0.0143 & 0.924 \\
       \\
        &                              & $\alpha_{02}$  & 2.611	    &0.401	&6.972	 &0.948  & &  --    &   --   &   --    &   --   \\
        &                              &$\alpha_{03}$   & --1.43	&1.190	&3.461	 &0.937  & &  --    &   --   &   --    &   --   \\
        & Regression Coefficient to model & $\alpha_{12}$  &1.410	    &1.191	&3.404	 &0.951  & &  --    &   --   &   --    &   --   \\
        & Recall probability for TTP &$\alpha_{13}$   &--1.743	&1.200	&4.467	 &0.949  & &  --    &   --   &   --    &   --   \\
        &                              &$\alpha_{1}$   & --1.330	&1.012	&2.789	 &0.943  & &  --    &   --   &   --    &   --   \\
        &                              &$\alpha_{2}$   & 1.067	    &1.052	&2.245   &0.961  & &  --    &   --   &   --    &   --   \\
        \\
        &  Probability of `very sure' recall  &                & 0.380	    &0.371	&0.2813	 &0.967  & &  --    &   --   &   --    &   --   \\

\noalign{\hrule height 1pt}
\end{tabular}
}
}
\vspace*{6pt}
\end{table}

\vspace*{6pt}In both Scenarios, it is found that the bias and the standard deviation (and consequently the MSE) of the Proposed Recall MLE is less than those of the other estimator. The bias of some parameters are large when the sample size is small but the performance of proposed MLE improves by increasing the sample size. Further, we found the coverage probability for the estimates are close to nominal 95\% and becomes even closer as the sample size increase.
As Table~\ref{t:tabletwo} shows even when the probability of certainty is considered to be equal, the improvement is found in the result of proposed MLE that accounts for the recall surety.
\section{A Real Data Analysis}
\label{dataAnalysis}
\subsection{Analysis of Upstate KIDS data}  \label{analysis}
We have analyzed the Upstate KIDS data set described in Section~\ref{intro} using our proposed models and methods. Various studies have assessed the association between time to pregnancy and soci-economic factors \citep{b1,b6,b3}. In our analysis, we considered four
socioeconomic variables: a continuous variable representing mothers' age, two categorical variables representing mother's education and her body mass index (BMI) respectively and one binary variable indicating mothers' smoking status.
We used the model proposed in Section~\ref{model}, discrete proportional hazards for time to pregnancy, the logistic model for planned pregnancy intention and the multinomial logistic model for the certainty attached to mothers' recalled TTP, $\pi_{\eta}^{(1)}, \pi_{\eta}^{(2)}, \pi_{\eta}^{(3)}$ and $\pi_{\eta}^{(4)}$, which are functions of `gap time from pregnancy to the observation time' and common determinants for TTP, as in previous section. We have considered the two different likelihood mentioned in Section~\ref{simulation} for estimating the parameters. Table~\ref{t:tablethree} gives a summary of the findings.

 \vskip-.05in
\begin{table}
  \caption{\label{t:tablethree}Estimate and standard error of coefficients from different method for Upstate KIDS data}
\hskip-.5in
 \centering
    \makebox[\textwidth][c]{
\scalebox{.8}{
\begin{tabular}{l c  c}
     \noalign{\hrule height 1pt}
                         & Proposed Recall MLE                 & Norecall MLE                 \\
\\
  Variable            &Estimate (SE)  &Estimate (SE)   \\
         \noalign{\hrule height 1pt}
                                     &Regression coefficient to model planned/ unplanned pregnancy ($\gamma$)& \\
 \cline{2-3}
 Maternal age                            & 0.031 (0.0007)  &  0.034 (0.0054)    \\
 Education (some college)       & 0.110	 (0.0327)  &  0.173  (0.0951)    \\
 Education (college \& above)   & 0.691  (0.0334)  &  0.742  (0.0375)    \\
 BMI (overweight)               & 0.022  (0.0171) &  0.017  (0.0226)    \\
 BMI (obese)                    & --0.320 (0.0147) &  --0.353 (0.0315)    \\
 Smoking status                 & --0.750 (0.0251) &  --0.653  (0.0282)    \\
 \noalign{\hrule height 0.5pt}
                      &    Regression coefficient for planned pregnancy TTP model ($\beta$)& \\
  \cline{2-3}
  Maternal age                           & --0.088 (0.0042)  &  --0.091 (0.0594)    \\
  Education (some college)      & 0.326  (0.0426)  &  0.362  (0.2056)    \\
  Education (college \& above)  & 0.447  (0.0164)  &  0.412  (0.0168)    \\
  BMI (overweight)              & 0.061  (0.1080)  &  0.064  (0.2774)    \\
  BMI (obese)                   & --0.028 (0.1303) &  --0.030 (0.2976)    \\
 Smoking status                 & --0.312 (0.0040)  &  --0.321  (0.0067)    \\
         \noalign{\hrule height 0.5pt}
                         &Regression coefficient for unplanned pregnancy TTP model ($\phi$)&\\
  \cline{2-3}
  Maternal age                           & --0.025 (0.0620)  &  --0.021 (0.124)    \\
  Education (some college)      & --0.312 (0.7901)  &  --0.380 (0.924)    \\
  Education (college \& above)  & --1.516 (0.9401)  &  --1.577 (1.1045)    \\
  BMI (overweight)              & 0.968  (0.5780)  &  0.979  (0.8258)    \\
  BMI (obese)                   & --0.036 (0.4066)  &  --0.034 (0.7598)    \\
 Smoking status                 & --2.134 (1.0080)  &  --2.384  (1.0578)    \\
        \noalign{\hrule height 0.5pt}
                        &Regression coefficient for recall probability ($\alpha$)&\\
  \cline{2-3}
  Maternal age                           & 0.044 (0.0091)  &  --    \\
  Education (some college)      & 0.153 (0.0279)  &  --    \\
  Education (college \& above)  & 0.741 (0.0349)  &  --    \\
  BMI (overweight)              & 0.019  (0.0160)  &  --    \\
  BMI (obese)                   & 0.253 (0.174)  &  --    \\
 Smoking status                 & --0.816 (0.0242)  &  --   \\
        \noalign{\hrule height 0.5pt}

 \noalign{\hrule height 1pt}
\end{tabular}
}
}
\vspace*{6pt}
\end{table}

 We find in Table~\ref{t:tablethree}, the estimated value of parameters in two models are close to each other while the standard errors of the Proposed Recall MLE are smaller than the corresponding standard errors of the other estimator. It is seen that mothers who are older or obese and those who are smoker took longer time to get pregnant, i.e. \, longer TTP.

\begin{figure}
\centering
 \includegraphics[height=5in,width=4in,angle=-90]{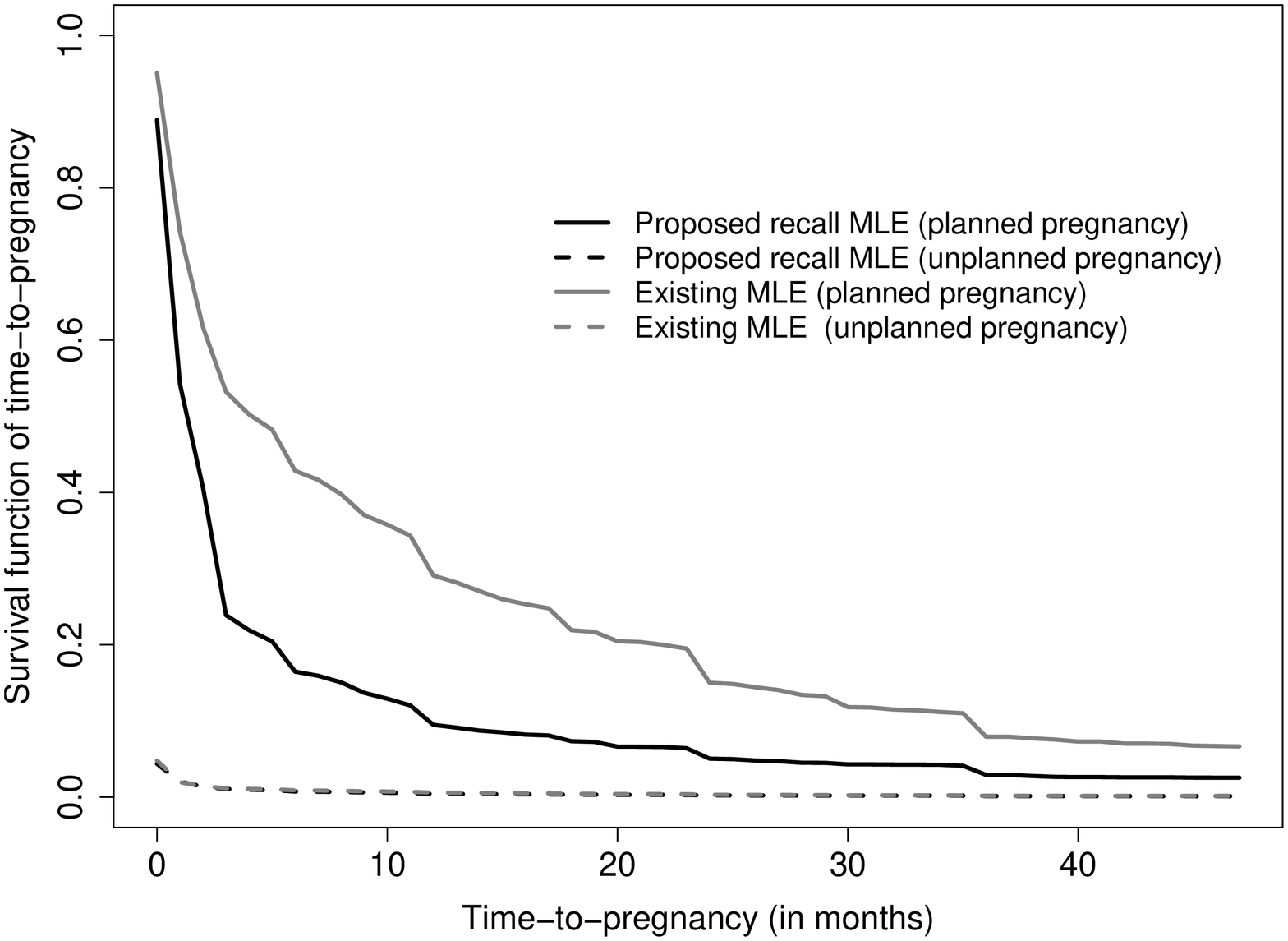}
\caption[]{Estimated survival functions of TTP based on Proposed Recall MLE, Norecall probability MLE and Existing MLE for the Upstate KIDS data.} \label{fig_2}
\end{figure}

Note that the model proposed by Sundaram et~al. (2012)\cite{b19}, can not account for the uncertainty information attached to recalled TTP. Here, the estimator of their model is referred as Existing MLE. As a comparison, the survival functions estimated from the Proposed Recall MLE and the Existing MLE for planned and unplanned pregnancy are shown in Figure~\ref{fig_2}. It is clear that  utilizing sampling weights and recall probabilities in the proposed MLE significantly change the survival function of TTP.

\subsection{Prediction for Upstate KIDS data}  \label{predict}
In this section, we predict survival function of TTP for Upstate KIDS data. This study contains 1793 nonresponding mothers and for whom TTP is unavailable of which 1694 reported unplanned and 99 planned pregnancies. We have used the Proposed Recall MLE and have estimated the $\hat\pi_i$ for $i=1,2,\ldots,1793$ of those individuals in Upstate KIDS data who did not answer the TTP question.

\begin{figure}
\centering
 \includegraphics[height=5in,width=4in,angle=-90]{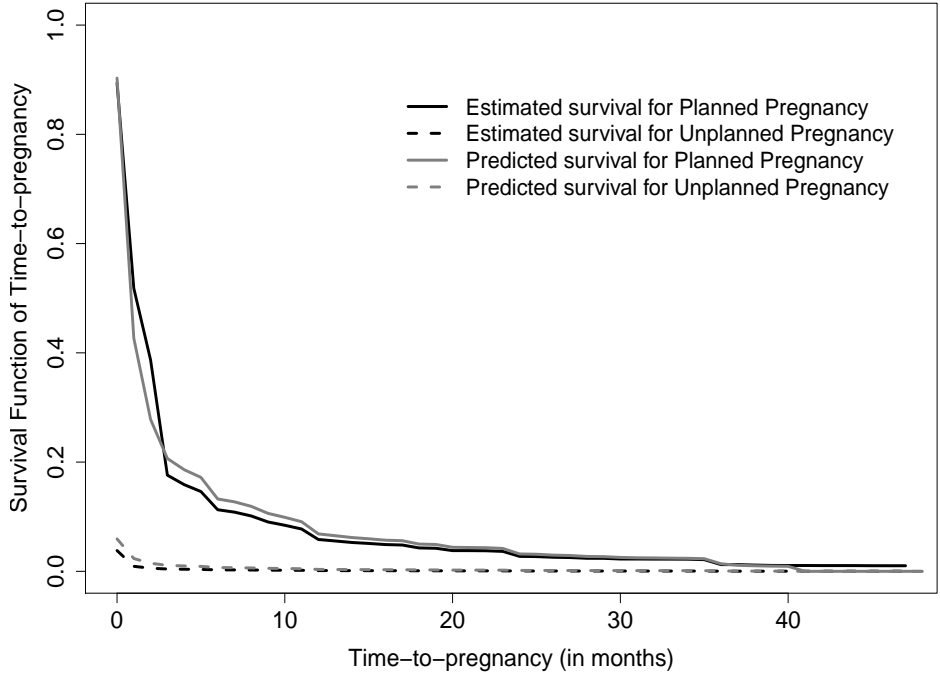}
\caption[]{Estimated and predicted survival functions for the the Upstate KIDS data.} \label{fig_3}
\end{figure}

\begin{figure}
\centering
 \includegraphics[height=6in,width=5in,angle=-90]{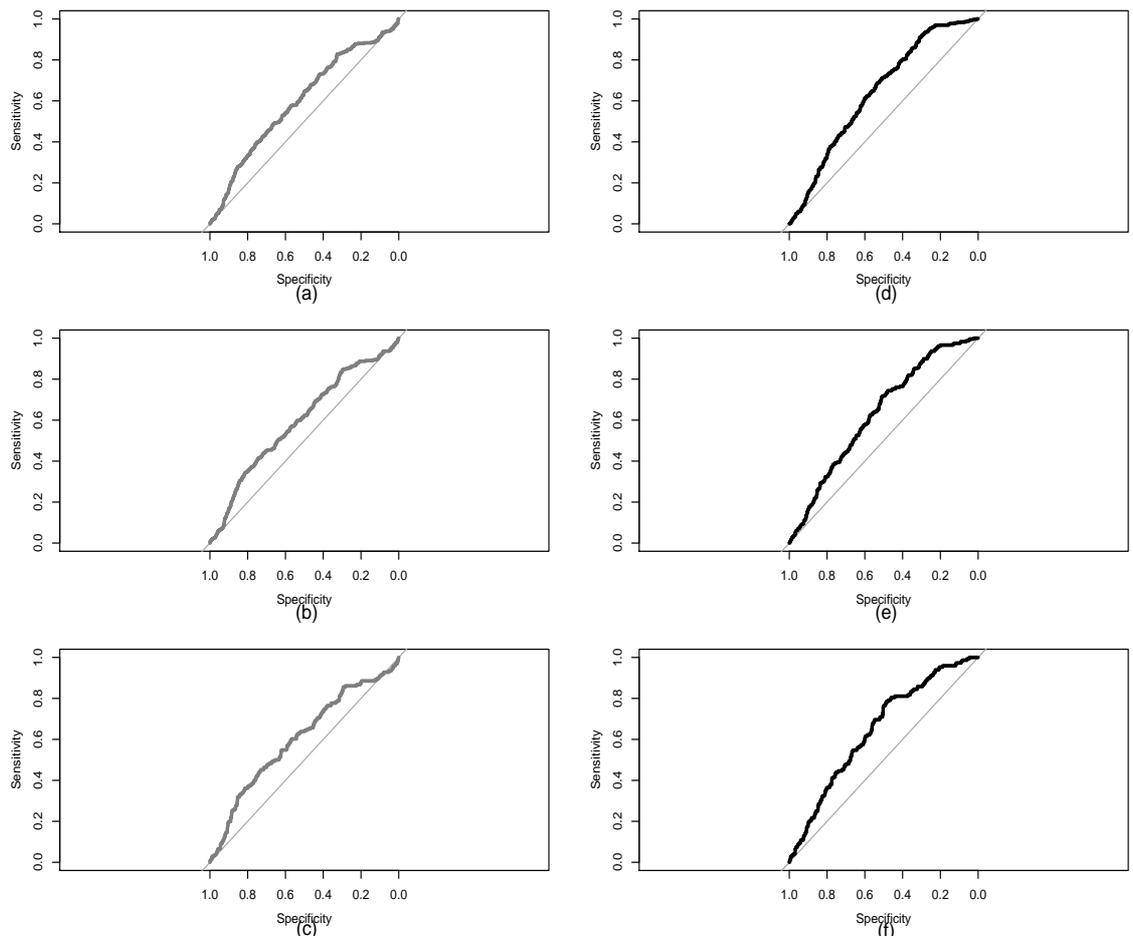}
\caption[]{ROC curve of classifying $I(T_i>t_0)$ for $t_0=6$ in (a) and (d), $t_0=12$ in (b) and (e), $t_0=20$ in (c) and (f), for two steps prediction using Proposed Recall MLE and prediction given planned/unplanned pregnancy, respectively.} \label{fig_4}
\end{figure}

As an example, the estimates of the survival of $S_i(u|\theta)$ has been calculated for two randomly selected mothers with planned and unplanned pregnancy using the Monte Carlo empirical Bayes estimation sampling scheme detailed in Section~(\ref{prediction}).
Figure~\ref{fig_3} display the prediction of survival function for these two individuals along with estimated survival function for planned and unplanned pregnancy using the proposed MLE.
To check the adequacy of prediction, we consider individuals with complete information. We have assumed 1/3 of them didn't provide the answer for TTP, considering planned/unplanned pregnancy and have estimated the parameter of the model based on the rest 2/3 of complete responses. Then we used the prediction method proposed in Section~\ref{prediction}, to predict the survival function for those individuals who hypothetically didn't provide their TTP. The performance of this prediction is quantified by receiver operating characteristic (ROC) curve and area under curve (AUC).
In this example, we were interested in predicting the probabilities of $(TTP>t_0)$ for $t_0=6,12,20$. That is, we classify $I(T_i>t_0)$ by the survivor $\pi_i(t_0|0)$ for $t_0=6,12,20$. To measure the classification rate we empirically estimate the sensitivity $P\{\hat\pi_i(t_0|0)>c|T_i>t_0\}$ and specificity $P\{\hat\pi_i(t_0|0)\leq c|T_i\leq t_0\}$ for $t_0=6,12,20$. These classification measures are displayed in an ROC curve in Figure~\ref{fig_4}~ (a) for $t_0=6$, (b) for $t_0=12$ and (c) for $t_0=20$, for all $c\in [0,1]$.
Also, we have repeated the prediction of probabilities of $(TTP>t_0)$ for $t_0=6,12,20$ when the planned/unplanned pregnancy is given. The results are shown in Figure~\ref{fig_4}~ (d) for $t_0=6$, (e) for $t_0=12$ and (f) for $t_0=20$, for all $c\in [0,1]$.
The AUC and their 95\% confidence interval correspond to each plot is as follow.
\begin{enumerate}
\item[(a)] For $t_0=6$, AUC=0.601 and 95\% C.I.=(0.561,0.643),
\item[(b)] For $t_0=12$, AUC=0.605 and 95\% C.I.=(0.558,0.646),
\item[(c)] For $t_0=20$, AUC=0.610 and 95\% C.I.=(0.565,0.651),
\item[(d)] For $t_0=6$ (when plan/unplan pregnancy is given), AUC=0.640 and 95\% C.I.=(0.616,0.681),
\item[(e)] For $t_0=12$ (when plan/unplan pregnancy is given), AUC=0.647 and 95\% C.I.=(0.605,0.701),
\item[(f)] For $t_0=20$ (when plan/unplan pregnancy is given), AUC=0.650 and 95\% C.I.=(0.602,0.704).

\end{enumerate}

\begin{figure}
\centering
 \includegraphics[height=6in,width=5in,angle=-90]{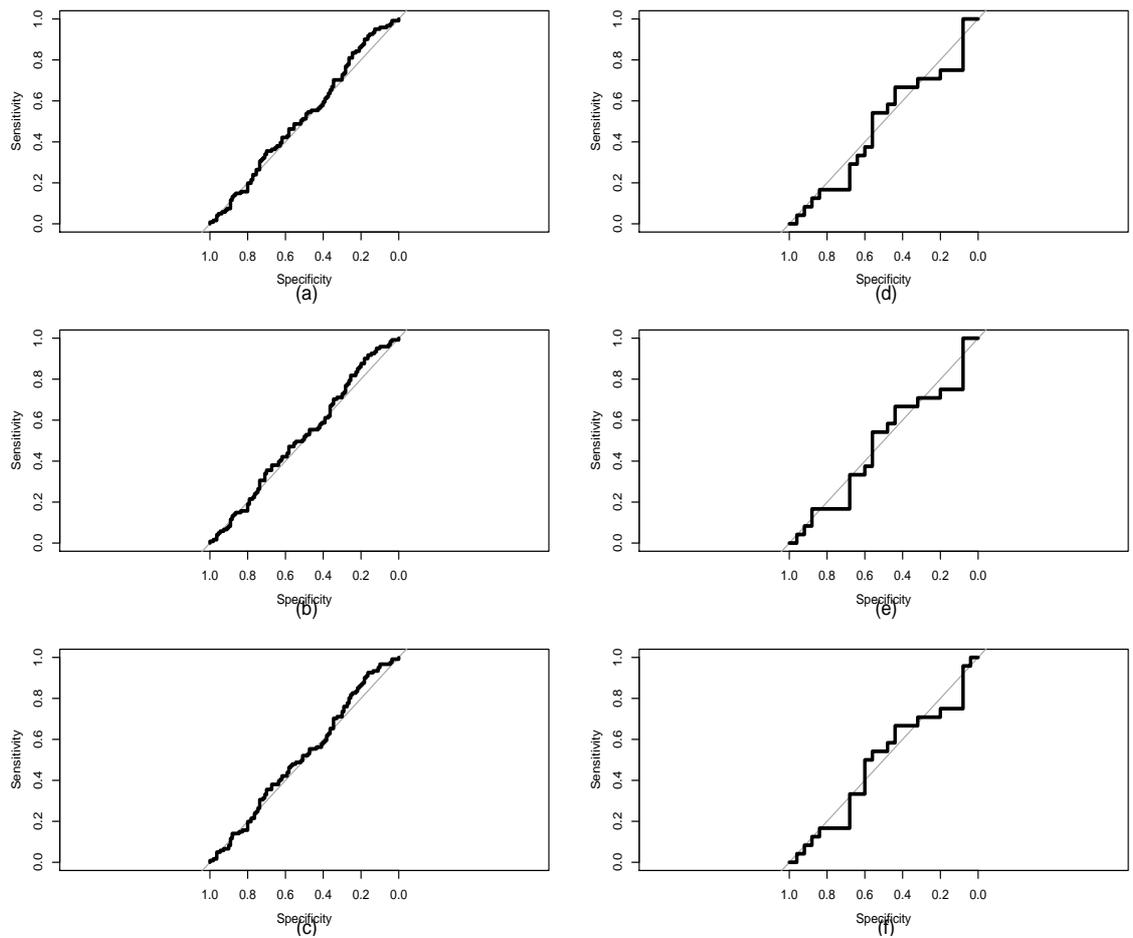}
\caption[]{ROC curve of classifying $I(T_i>t_0)$ for $t_0=6$ in (a) and (d), $t_0=12$ in (b) and (e), $t_0=20$ in (c) and (f), for two steps prediction using Existing MLE and prediction given planned/unplanned pregnancy, respectively.} \label{fig_5}
\end{figure}

As a comparison, we repeated the prediction procedure by using the Existing MLE. Figure~\ref{fig_5} shows similar ROC curves for cases (a)--(e) mentioned above using Existing MLE for the prediction without considering sampling weight and uncertainty information associated with retrospective TTP. The AUC and their 95\% confidence intervals correspond to each plot in Figure~\ref{fig_5} is as follow.
\begin{enumerate}
\item[(a)] For $t_0=6$, AUC=0.520 and 95\% C.I.=(0.501,0.638),
\item[(b)] For $t_0=12$, AUC=0.521 and 95\% C.I.=(0.502,0.616),
\item[(c)] For $t_0=20$, AUC=0.521 and 95\% C.I.=(0.505,0.640),
\item[(d)] For $t_0=6$ (when plan/unplan pregnancy is given), AUC=0.478 and 95\% C.I.=(0.416,0.638),
\item[(e)] For $t_0=12$ (when plan/unplan pregnancy is given), AUC=0.482 and 95\% C.I.=(0.425,0.633),
\item[(f)] For $t_0=20$ (when plan/unplan pregnancy is given), AUC=0.483 and 95\% C.I.=(0.368,0.635).

\end{enumerate}

The improvement of the prediction and increase of AUC from the proposed MLE are clearly seen by comparing Figures~\ref{fig_4} and~\ref{fig_5}.


\section{Concluding Remarks}
\label{concluding}
In this paper, we have offered a realistic model and method
for estimating the discrete survival time distribution based on recalled data. Modeling the
uncertainty attached with the recall of time is a unique feature of our proposed model for the retrospectively measured survival time. We find our prediction for infertility, i.e., TTP $ > 12$ or $ > 20$ is reasonable based on proposed model using very basic socio-demographic factors as covariates. We believe including prior history of maternal reproductive history and overall health as well as paternal factors, consistent with fecundity being a couple based outcome in the proposed model would considerably improve the predictive ability of the proposed approach and increases the AUC. Our simulation results show that using woman's certainty of TTP data improves the estimators' performance. This may be something that researchers designing surveys requiring retrospective information should be encouraged to ask given the errors associated with retrospective time-to-event.

Grouping of the uncertainly recalled event may reduce the recall error to some extent. If one adopts this solution, the method presented here is a viable method of analysis.
Skinner and Humphreys (1999)\cite{b18}, while working with data without instances of missing, has modeled erroneously recalled time-to-event as $t'_i=t_ik_i$, where $t_i$ is the correct time-to-event and $k_i$ is a multiplicative error of recall that is independent of $t_i$. They have used a mixed-effects regression model to account for $k_i$. One may investigate whether a similar adjustment in the likelihood~\eqref{ourM}, improves the analysis.

There can be an alternative approach for modeling this kind of data, through an underlying distribution ($F$)
for the time to occurrence of the event of interest, and
another distribution (say, $G$) for the time from that occurrence to
the forgetting of the exact time. The latter may in fact be a
sub-distribution function, with some mass at infinity. In this
formulation there would be two cases for individual $i$:
only the first event has occurred and both events have occurred. Depending on the relation between the two events and the observation time, one can model the joint distribution of these events and based on that, estimate the marginal distribution of first event which is the goal in this study.

\section*{Acknowledgements}
 This research was supported by the Intramural Research Program of the {\it Eunice Kennedy Shriver} National Institute of Child Health and Human Development (NICHD; contracts \#HHSN275201200005C, \#HHSN267200700019C).
 This work was completed while the first author was a `Visiting  Fellow' with Dr. Sundaram. The authors also acknowledge that this study utilized the high performance computational capabilities of the Biowulf Linux cluster at the National Institutes of Health, Bethesda, Maryland (http://biowulf.nih.gov).
\newpage


\newpage
\renewcommand\baselinestretch{1.3}
\normalsize

\end{document}